\documentclass[letterpaper, 10pt, conference]{ieeeconf}

\usepackage{decar-common}
\usepackage{decar-dynamics}
\usepackage{decar-lie}
\usepackage{decar-post}

\makeatletter
\AtBeginDocument{
  \check@mathfonts
}
\makeatother

\usepackage{flushend}

\def\BibTeX{{\rm B\kern-.05em{\sc i\kern-.025em b}\kern-.08em
    T\kern-.1667em\lower.7ex\hbox{E}\kern-.125emX}}

\title{\LARGE \bf
Design of Input-Output Observers for a Population of Systems
with Bounded Frequency-Domain Variation using $DK$-iteration
}

\author{Timothy Everett Adams and James Richard Forbes$^{1}$
\thanks{This work was supported by the NSERC Discovery Grant Program and NSERC
  Alliance program in collaboration with Denso.}%
\thanks{$^{1}$The authors are with the Department of Mechanical Engineering,
McGill University, 817 Sherbrooke St. W., Montreal, QC H3A 0C3, Canada. The
corresponding author is {\tt\small timothy.adams@mail.mcgill.ca}.}%
}

\begin{document}

%
%
%
%
%
%
%
\def \myJournal {IEEE Control Systems Letters}
\def \myDoi {10.1109/LCSYS.2025.3638270}
\def \myPaperSiteName {IEEE Xplore}
\def \myPaperSiteLink {https://ieeexplore.ieee.org/document/11271285}
\def \myYear {2025}
\def \myPaperCitation{T. E. Adams and J. Richard Forbes, ``Design of Input-Output Observers for a Population of Systems With Bounded Frequency-Domain Variation Using DK-Iteration,'' in \textit{IEEE Control Systems Letters}, vol. 9, pp. 2645-2650, 2025.}


\begin{figure*}[t]

\thispagestyle{empty}
\begin{center}
\begin{minipage}{6in}
\centering
This paper has been accepted for publication in \emph{\myJournal}. 
\vspace{1em}

This is the author's version of an article that has, or will be, published in this journal or conference. Changes were, or will be, made to this version by the publisher prior to publication.
\vspace{2em}

\begin{tabular}{rl}
DOI: & \myDoi\\
\myPaperSiteName: & \texttt{\myPaperSiteLink}
\end{tabular}

\vspace{2em}
Please cite this paper as:

\myPaperCitation

\vspace{15cm}
\copyright \myYear \hspace{4pt}IEEE. Personal use of this material is permitted. Permission from IEEE must be obtained for all other uses, in any current or future media, including reprinting/republishing this material for advertising or promotional purposes, creating new collective works, for resale or redistribution to servers or lists, or reuse of any copyrighted component of this work in other works.

\end{minipage}
\end{center}
\end{figure*}
\newpage
\clearpage
\pagenumbering{arabic}

\pagestyle{empty}
\maketitle
\thispagestyle{empty}

\begin{abstract}
    This paper proposes a linear input-output observer design methodology for a
    population of systems in which each observer uses knowledge of the linear
    time-invariant dynamics of the particular device.
    Observers are typically composed of a known model of the system and a
    correction mechanism to produce an estimate of the state.
    The proposed design procedure characterizes the population variation in the
    frequency domain and synthesizes a single robust correction filter.
    The correction filter guarantees a level of estimation performance for all
    systems compatible with the uncertainty characterization.
    This is accomplished by posing a robust performance problem using the
    observer error dynamics and solving it using $DK$-iteration.
    The design procedure is experimentally demonstrated on a flexible joint
    robotic manipulator with varied joint stiffnesses.
    It is shown that the proposed robust correction filter achieves comparable
    estimation performance to a method using a correction gain tailored toward
    each joint stiffness configuration.
\end{abstract}


\section{Introduction}
\label{sec:intro}

In many applications, knowledge of the state of a system is
required. 
For example, the state is required to perform state-feedback control
\cite{skogestadMultivariableFeedbackControl2005},
\cite{williamsLinearStateSpaceControl2007}, fault detection
\cite{miljkovićFaultDetectionMethods2011}, and system monitoring
\cite{zhaoPowerSystemRealTime2016}.
The full state of a system is typically not measured due to limitations on the
number and placement of sensors.
Observers are used to estimate the state of the system using a model of the
system dynamics, as well as knowledge of the inputs and outputs applied to the
system.
Typically, an observer predicts the state to be estimated using the system
dynamics model and the known input augmented with a correction term driven by
the innovation, which is the error between the measured and estimated system
output.

There are many methods for synthesizing observers, also often referred to as
state estimators.
The most common observers realize the correction step via a multiplication of
the innovation by a constant gain.
In the Kalman filter, the constant gain is designed to minimize the
mean-squared estimate error subject to white noise disturbances
\cite{andersonOptimalFiltering1979}, \cite{simonOptimalStateEstimation2006}.
In $\mc{H}_{2}$ and $\mc{H}_{\infty}$ observers, the constant gain minimizes
the gain from disturbances to state estimate errors measured via the
$\mc{H}_{2}$ and $\mc{H}_{\infty}$ norms, respectively
\cite{simonOptimalStateEstimation2006}, \cite{greenLinearRobustControl2012}.
More general observer structures such as the input-output and dynamical
observers substitute the correction gain for a dynamical filter, that is to say
a transfer matrix, which allows for additional degrees of freedom in the design
\cite{marquezFrequencyDomainApproach2003},
\cite{marquezRobustStateObserver2005}.
Often the constant gain and dynamic filter are referred to as the correction
gain and correction filter, respectively.

\begin{figure}[t]
  \centering
  \begin{subfigure}[t]{0.47\linewidth}
    \centering
    \includegraphics[width=\textwidth]{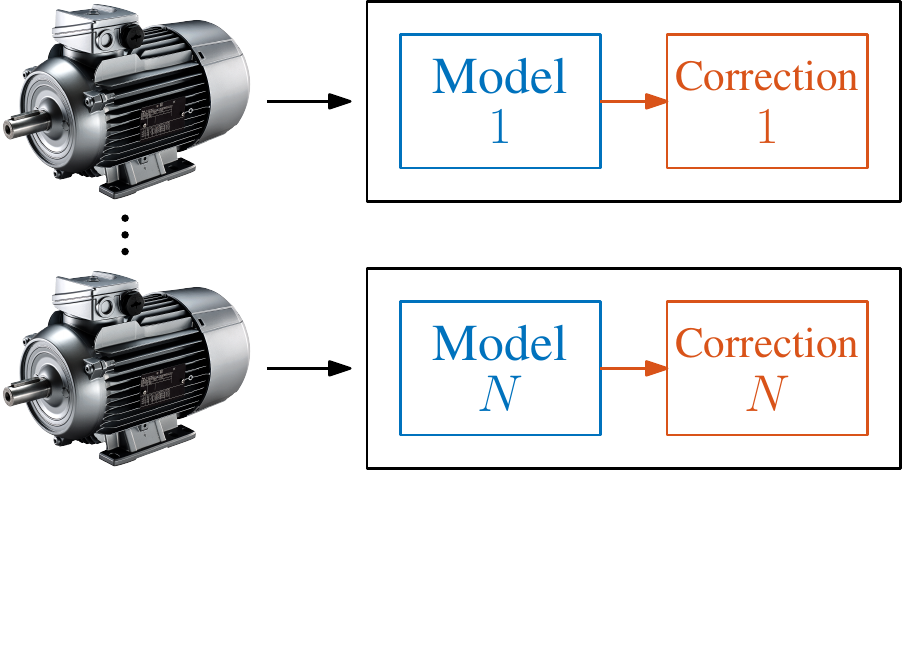}
    \caption{A correction step is designed for, and used with each of its
    respective system models to form an observer.}
    \label{fig:overview_diagram_tailored}
  \end{subfigure}
  \hfill
  \begin{subfigure}[t]{0.47\linewidth}
    \centering
    \includegraphics[width=\textwidth]{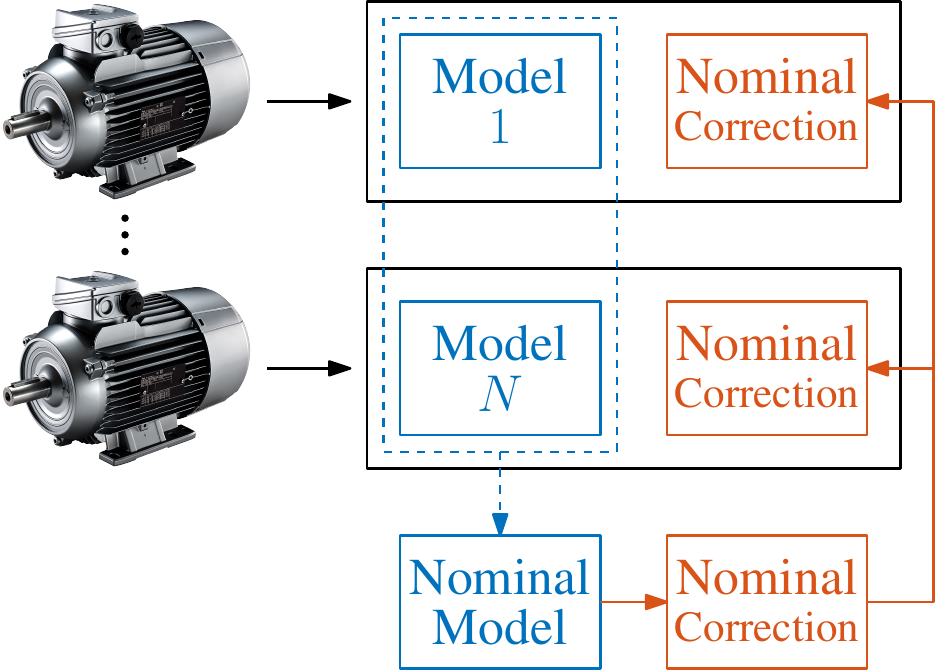}
    \caption{One correction step is designed using a nominal model, and used
    with each system model to form an observer.}
    \label{fig:overview_diagram_nominal}
  \end{subfigure}
  \hfill
  \begin{subfigure}[t]{0.47\linewidth}
    \centering
    \includegraphics[width=\textwidth]{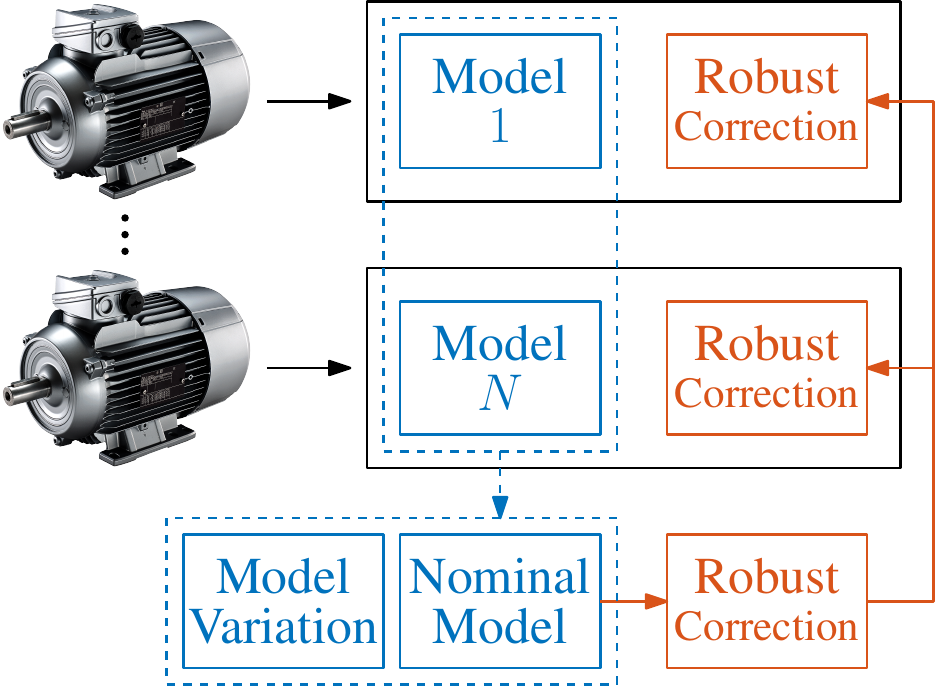}
    \caption{One correction step is designed using a nominal model and
    uncertainty characterization, and used with each system model to form an
    observer.}
    \label{fig:overview_diagram_robust}
  \end{subfigure}
  \caption{Diagram of observer design methodologies for a population of
  electric motors with manufacturing variation.}
  \vspace{-15pt}
  \label{fig:overview_diagram}
\end{figure}

In many cases, a set of observers must be designed for a batch of varied
devices with manufacturing variation, or that have dynamics that change slowly
over time due to wear and tear.
In these cases, the design problem is challenging as the variation in the
system dynamics may degrade the estimation performance.
A model of each device can be acquired through system identification (ID)
\cite{ljungSystemIdentificationTheory1999}.
For example, a system ID procedure can be applied to devices from an assembly
line, resulting in a data-driven model of each system.
The use of tailored data-driven models in the observers improves the estimation
accuracy as it exploits the particularities of each system.
Therefore, a methodology for designing observers that exploit a system ID model
of each device in a population is desirable.

Current observer synthesis methods only allow for the design of a correction
gain or filter from a single model.
This results in design methodologies that do not scale well with the number of
devices or lack estimation accuracy guarantees across the population.
These limitations are illustrated in an example in which $N$ electric
motors with varied dynamics due to manufacturing variation are outfitted with
observers.
All devices undergo a system ID procedure such that a model of each device is
available.
Fig.~\ref{fig:overview_diagram_tailored} shows a methodology in which
a correction mechanism is designed using each system model.
This approach provides good estimation accuracy as the correction is tailored
toward each system, but it becomes impractical when $N$ is large as a
correction step must be synthesized and tuned for each model.
Fig.~\ref{fig:overview_diagram_nominal} shows a procedure in which a single
correction step is designed from a nominal model extracted from the
population.
This method is practical as only one correction mechanism is designed, but it
does not yield estimation accuracy guarantees for devices in the population
different from the nominal model.
Therefore, there is a need for an observer design method for a population of
devices that scales with the number of systems and provides performance
guarantees across the population.

The contribution of this paper is a novel procedure for synthesizing a set of
input-output observers that exploit a data-driven model of each device in a
population.
The proposed method, depicted in Fig.~\ref{fig:overview_diagram_robust} and
detailed in Section~\ref{sec:methodology}, involves the synthesis of a single
robust correction filter from a nominal model and a characterization of the
population variation using linear robust control theory.
Then, the observers are formed by pairing the robust correction filter with
each data-driven device model.
Given that observers can be trivially formed once the correction filter is
designed, an observer can be made for any additional system that satisfies the
uncertainty characterization even if it wasn't from the initial population.
This results in a methodology that can scale with an arbitrarily large number
of devices provided they satisfy the uncertainty characterization.
Moreover, the robust correction filter provides an estimation accuracy
guarantee for all observers formed with models in the population through the
use of synthesis methods from robust control theory.
The proposed approach is experimentally demonstrated on the estimation of
angular positions in a planar flexible joint robotic manipulator subject to
variation in joint stiffnesses.

\section{Preliminaries}
\label{sec:preliminaries}

\subsection{Observer Design}
\label{subsec:state_observer_design}

This section outlines the input-output observer structure used in this paper
and the observer design problem.
Linear time-invariant (LTI) systems composed of a process and measurement model
are considered.
The process model $\mbf{G}(s) \in \cnums^{n_{x} \times n_{u}}$, defined as
\begin{align}
  \mbf{x}(s) = \mbf{G}(s) \mbf{u}(s),
  \label{eq:process_model}
\end{align}
relates the inputs $\mbf{u}(s)$ to the states $\mbf{x}(s)$.
The measurement model is defined as
\begin{align}
  \mbf{y}(s) = \mbf{C} \mbf{x}(s),
  \label{eq:measurement_model}
\end{align}
where $\mbf{C} \in \rnums^{n_{y} \times n_{x}}$ is a constant matrix that maps
the states $\mbf{x}(s)$ to the outputs $\mbf{y}(s)$.
It is assumed that the measurement model is a static mapping with no
feedthrough.

\begin{figure}[t]
  \vspace{5pt}
  \centering
  \includegraphics[scale=0.8]{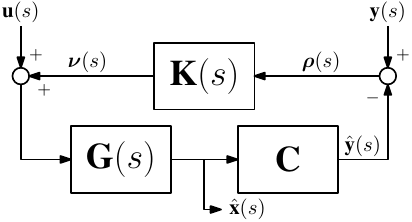}
  \caption{Input-output observer.}
  \vspace{-10pt}
  \label{fig:input_output_observer_diagram}
\end{figure}

An attractive observer structure is the input-output observer, depicted in
Fig.~\ref{fig:input_output_observer_diagram}, as it allows for
frequency-domain performance specifications and controller synthesis methods
such as $\mc{H}_{2}$- and $\mc{H}_{\infty}$-synthesis
\cite{marquezFrequencyDomainApproach2003},
\cite{marquezRobustStateObserver2005}.
The input-output observer is composed of the process model
\eqref{eq:process_model}, the measurement model \eqref{eq:measurement_model},
and the correction filter $\mbf{K}(s)$ defined as
\begin{align}
  \mbs{\nu}(s) = \mbf{K}(s) \mbs{\rho}(s).
  \label{eq:observer_correction_filter}
\end{align}
The filter $\mbf{K}(s) \in \cnums^{n_{u} \times n_{y}}$ maps the innovation
$\mbs{\rho}(s) = \mbf{y}(s) - \mbfhat{y}(s)$, which is the error between the
measured and estimated output, to the input correction $\mbs{\nu}(s)$.
The estimated output is $\mbfhat{y}(s) = \mbf{C} \mbfhat{x}(s)$ where
$\mbfhat{x}(s)$ is the estimated state.

\begin{figure}[t]
  \centering
  \vspace{5pt}
  \includegraphics[scale=0.8]{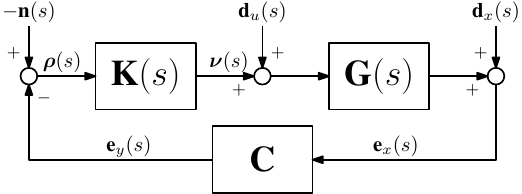}
  \caption{Input-output observer error dynamics.}
  \vspace{-15pt}
  \label{fig:observer_error_dynamics_diagram}
\end{figure}

The observer is designed by synthesizing $\mbf{K}(s)$ that stabilizes the
estimation error dynamics and mitigates the effect of disturbances on the errors.
The error dynamics, shown in Fig.~\ref{fig:observer_error_dynamics_diagram},
describe the effect of disturbances and/or noise at the input $\mbf{d}_{u}(s)$,
state $\mbf{d}_{x}(s)$, and output $\mbf{n}(s)$ on estimate errors at the state
$\mbf{e}_{x}(s)$ and output $\mbf{e}_{y}(s)$.
The error dynamics are in the form of a common control problem meaning that
standard synthesis methods can be used.

\subsection{Robust Control Theory}
\label{subsec:robust_control_theory}

This section outlines the robust control tools used to characterize
the variation within a population of devices and to synthesize a controller
that guarantees performance across all systems.
Dynamic frequency-dependent uncertainty models are used to characterize the
variation of a population of systems
\cite{skogestadMultivariableFeedbackControl2005}.
A system is described by the interconnection of a nominal LTI model
$\mbf{G}_0(s)$, a weight $\mbf{W}_{\Delta}(s)$, and a bounded complex
perturbation $\mbs{\Delta}(s)$ defined as any LTI system that satisfies
\begin{align}
  \norm{\mbs{\Delta}(s)}_{\infty} = \sup_{\omega} \bar{\sigma}(\mbs{\Delta}(j\omega)) \leq 1.
  \label{eq:delta_block}
\end{align}
For example, a multiplicative input uncertainty model describes a dynamics
model in a set as $\mbf{G}(s) = \mbf{G}_{0}(s) (\eye + \mbf{E}_{I}(s))$ where
$\mbf{E}_{I}(s) = \mbf{W}_{\Delta}(s) \mbs{\Delta}(s)$ is the residual.
There are other uncertainty models each suited to different applications
\cite{skogestadMultivariableFeedbackControl2005}.

\begin{figure}[t]
  \centering
  \vspace{5pt}
  \begin{subfigure}[b]{0.47\linewidth}
    \centering
    \includegraphics[width=\textwidth]{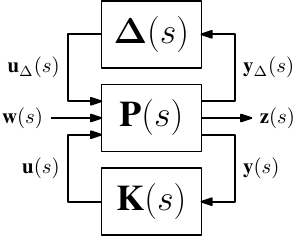}
    \caption{General structure for controller synthesis.}
    \label{fig:robust_control_diagram_general}
  \end{subfigure}
  \hfill
  \begin{subfigure}[b]{0.47\linewidth}
    \centering
    \includegraphics[width=\textwidth]{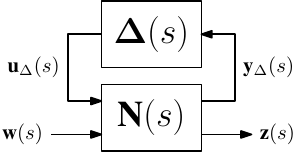}
    \caption{$\mbs{N}\mbs{\Delta}$-structure for robust performance analysis.}
    \label{fig:robust_control_diagram_n_delta}
  \end{subfigure}
  \caption{General robust control problem block diagram.}
  \vspace{-10pt}
  \label{fig:robust_control_diagrams}
\end{figure}

Fig.~\ref{fig:robust_control_diagram_general} depicts a block diagram of the
general robust control design problem.
The generalized plant $\mbf{P}(s)$ interconnects the controller $\mbf{K}(s)$
and perturbation $\mbs{\Delta}(s)$ with the system to be controlled along with
exogenous disturbances $\mbf{w}(s)$ and performance outputs $\mbf{z}(s)$ to be
minimized.
The objective is to design $\mbf{K}(s)$ that ensures the gain from $\mbf{w}(s)$
to $\mbf{z}(s)$ is below a given threshold for all perturbations
$\mbs{\Delta}(s)$ satisfying \eqref{eq:delta_block}, which is called the robust
performance problem.
In order to establish the robust performance condition, let the system
$\mbf{N}(s)$ shown in Fig.~\ref{fig:robust_control_diagram_n_delta} be
defined using a lower linear fractional transformation as
\cite{skogestadMultivariableFeedbackControl2005}
\begin{align}
  \mbf{N}(s) &= \mbf{F}_{\ell}(\mbf{P}(s), \mbf{K}(s)).
  \label{eq:n_system}
\end{align}
The robust performance condition is defined as
\begin{align}
  \mu_{\mbshat{\Delta}}(\mbf{N}(j\omega)) < 1, \quad \forall \omega \in [0, \infty),
  \label{eq:robust_performance_ssv}
\end{align}
where $\mu_{\mbshat{\Delta}}$ is the structured singular value (SSV) associated
with the perturbation $\mbshat{\Delta}(s)$
\cite{skogestadMultivariableFeedbackControl2005}.
The structured perturbation $\mbshat{\Delta}(s)$ is given by
\begin{align}
  \mbshat{\Delta}(s) = 
  \begin{bmatrix}
    \mbs{\Delta}(s) & \mbf{0} \\
    \mbf{0} & \mbs{\Delta}_{p}(s)
  \end{bmatrix},
  \label{eq:robust_performance_delta}
\end{align}
where $\mbs{\Delta}_{p}(s)$ is a fictitious perturbation associated with the
performance channel gain from $\mbf{w}(s)$ to $\mbf{z}(s)$.
The signals $\mbf{w}(s)$ and $\mbf{z}(s)$ must be appropriately normalized for
\eqref{eq:robust_performance_ssv} to hold any significance.
In practice, $\mu_{\mbshat{\Delta}}$ is not directly computed.
Instead, the upper bound
\begin{align}
  \mu_{\mbshat{\Delta}}(\mbf{N}(j\omega)) \leq \min_{\mbf{D}(j\omega) \in \mbc{D}_{\mbshat{\Delta}}} \bar{\sigma}(\mbf{D}(j\omega) \mbf{N}(j\omega) \mbf{D}(j\omega)\inv)
  \label{eq:ssv_upper_bound}
\end{align}
is used where $\mbc{D}_{\mbshat{\Delta}}$ is the set of matrices that commute
with $\mbshat{\Delta}(j\omega)$ such that $\mbf{D}(j\omega)
\mbs{\Delta}(j\omega) = \mbs{\Delta}(j\omega) \mbf{D}(j\omega)$
\cite{skogestadMultivariableFeedbackControl2005}.
The upper bound in \eqref{eq:ssv_upper_bound} is a generalized eigenvalue
problem that can be solved by a bisection on the upper bound with semidefinite
feasibility problems on $\mbf{D}(j\omega)$ at discrete frequencies $\omega$
\cite{caverlyLMIPropertiesApplications2024}.
Hence, an appropriate grid of frequencies must be selected to evaluate the
robust performance condition.

Currently, there are no direct methods to synthesize $\mu$-optimal controllers.
However, there is a method known as $DK$-iteration, which alternates between
$\mc{H}_{\infty}$-synthesis and $\mu$-analysis, that works well at achieving
robust performance in practice
\cite{skogestadMultivariableFeedbackControl2005}.
This method designs a controller that minimizes the upper bound on the SSV in
\eqref{eq:ssv_upper_bound} yielding the optimization problem
\begin{align}
  \min_{\mbf{K}(s)} \left( \min_{\mbf{D}(s) \in \mbc{D}_{\mbshat{\Delta}}} \norm{\mbf{D}(s) \mbf{N}(s) \mbf{D}(s)\inv}_{\infty} \right),
  \label{eq:dk_iteration_optimization}
\end{align}
which is solved by alternating the minimization of the objective function
between the controller $\mbf{K}(s)$ and the scaling matrix $\mbf{D}(s)$
\cite{skogestadMultivariableFeedbackControl2005}.
Given that the scaling matrices $\mbf{D}(j\omega)$ can only be computed at
discrete frequencies, a stable and minimum-phase transfer matrix must be fit to
the magnitude response to obtain $\mbf{D}(s)$.
This results in the following procedure.
\begin{packed_enum}
  \setcounter{enumi}{-1}
  \item Select an appropriate discrete grid of frequencies.
  \item Synthesize an $\mc{H}_{\infty}$-optimal controller $\mbf{K}(s)$ for the
    system $\mbf{D}(s) \mbf{N}(s) \mbf{D}(s)\inv$.
  \item Compute the upper bound of $\mu_{\mbshat{\Delta}}(\mbf{N}(j\omega))$
    and scaling matrix $\mbf{D}(j\omega)$ at each frequency. Stop
    if robust performance is achieved, otherwise continue.
  \item Fit a stable and non-minimum phase transfer matrix $\mbf{D}(s)$ to the
    magnitude of $\mbf{D}(j\omega)$ and return to step 1.
\end{packed_enum}

\section{Methodology}
\label{sec:methodology}

The proposed observer design method poses and solves a robust performance
problem for the input-output observer error dynamics using $DK$-iteration.
The methodology is as follows.
\begin{packed_enum}
  \item Identify an LTI model for each device in a population using a
    standardized system ID procedure.
  \item Characterize the variation of the population using a dynamic
    uncertainty model.
  \item Design performance weights and form a generalized plant for the
      input-output observer error dynamics.
  \item Synthesize an observer correction filter using $DK$-iteration that
    achieves robust performance.
  \item Form an input-output observer for each device by combining the robust
    correction filter with the respective system's LTI model.
\end{packed_enum}

The robust performance criterion implies that the gain from the disturbances
and noise to the estimation error is below a threshold for all devices
in the population.
Given that the performance is attained for the population using one correction
filter, the estimation accuracy of each observer can be slightly worse than if
a tailored correction method was used due to the trade-off between robustness
and performance.
However, the proposed method is favorable as only a single correction
filter must be designed for numerous devices.

The number of devices in the population that satisfy the performance level with
the robust correction filter may be arbitrarily large.
The only factor that dictates if a device will satisfy the performance
threshold is whether it is compliant with the uncertainty characterization.
Therefore, an observer can be formed for a new device even after the
uncertainty characterization has been completed provided it is
within the uncertainty bounds.
This highlights the importance of the uncertainty characterization accuracy as
it is the determining factor in whether a device in the population will achieve
the desired estimation accuracy.
The uncertainty characterization can be extended to nonlinear systems by
including linearized models about different equilibrium points in the
population.

\section{Experimental Results}
\label{sec:results}

\begin{figure}[t]
  \centering
  \vspace{10pt}
  \includegraphics[width=0.9\linewidth]{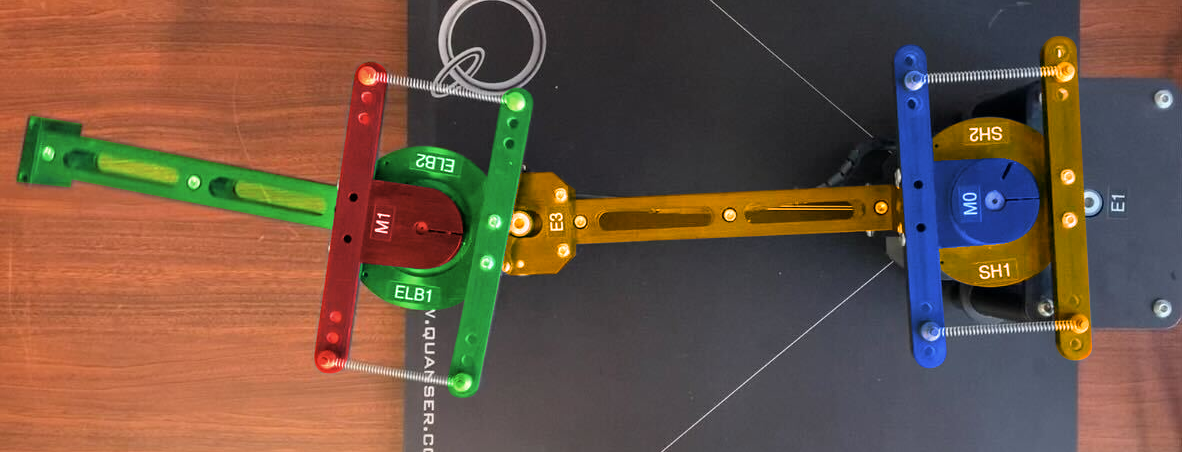}
  \caption{Picture of Quanser \textit{2 DOF Serial Flexible Joint} with
  colorized rigid bodies (blue: shoulder motor hub; orange: shoulder link; red:
  elbow motor hub; green: elbow link).}
  \label{fig:flexible_joint_robot}
\end{figure}

The state observer design methodology is demonstrated on the Quanser \textit{2
DOF Serial Flexible Joint} shown in Fig.~\ref{fig:flexible_joint_robot}.
The flexible joint manipulator has a planar shoulder and elbow joint.
Each joint is composed of a hub rigidly connected to the motor shaft, and a
link connected to the motor hub via a pair of springs to exaggerate the
effect of gearbox compliance.
The radial position and stiffness of the springs is varied to simulate the
effect of manufacturing variation in gearbox stiffness.
Each body has an encoder that measures its angular position.

The estimation task is to infer the angular positions of the motor hubs
and link elastic deflections using the motor currents and motor hub encoder
measurements for a set of $4$ different joint stiffness configurations.
The dynamics of the manipulator are nonlinear whereas the observer framework
proposed in Section~\ref{sec:methodology} only applies to linear systems.
Therefore, the estimation is performed in a linear regime of small
perturbations about the equilibrium point in which both links are colinear and
stationary.
The code required to reproduce the results of this paper is available at
\url{https://github.com/decargroup/input_output_observer_population}.

Table~\ref{tab:fjr_system_variables} shows the system variables for the
flexible joint manipulator.
The inputs, states, and outputs of the system are defined as
\begin{align}
  \mbf{u}(s) &= 
  \begin{bmatrix}
    i_{1}(s) & i_{2}(s)
  \end{bmatrix}^{\trans},
  \label{eq:inputs} \\
  \mbf{x}(s) &= 
  \begin{bmatrix}
    \theta_{1}(s) & \alpha_{1}(s) & \theta_{2}(s) & \alpha_{2}(s) 
  \end{bmatrix}^{\trans},
  \label{eq:states} \\
  \mbf{y}(s) &= 
  \begin{bmatrix}
    \theta_{1}(s) & \theta_{2}(s)
  \end{bmatrix}^{\trans},
  \label{eq:outputs}
\end{align}
respectively.

\begin{table}[t]
  \centering
  \vspace{5pt}
  \caption{Flexible joint manipulator system variables.}
  \begin{tabular}{l c c}
    \hline \hline
    Parameter & Unit & Symbol \\
    \hline
    Shoulder motor current & A & $i_{1}$ \\
    Elbow motor current & A & $i_{2}$ \\
    Shoulder motor hub angular position & deg & $\theta_{1}$ \\
    Shoulder link angular deflection & deg & $\alpha_{1}$ \\
    Elbow motor hub angular position & deg & $\theta_{2}$ \\
    Elbow link angular deflection & deg & $\alpha_{2}$ \\
    \hline \hline
  \end{tabular}
  \vspace{-10pt}
  \label{tab:fjr_system_variables}
\end{table}

\begin{figure}[t]
  \centering
  \includegraphics[width=1.0\linewidth]{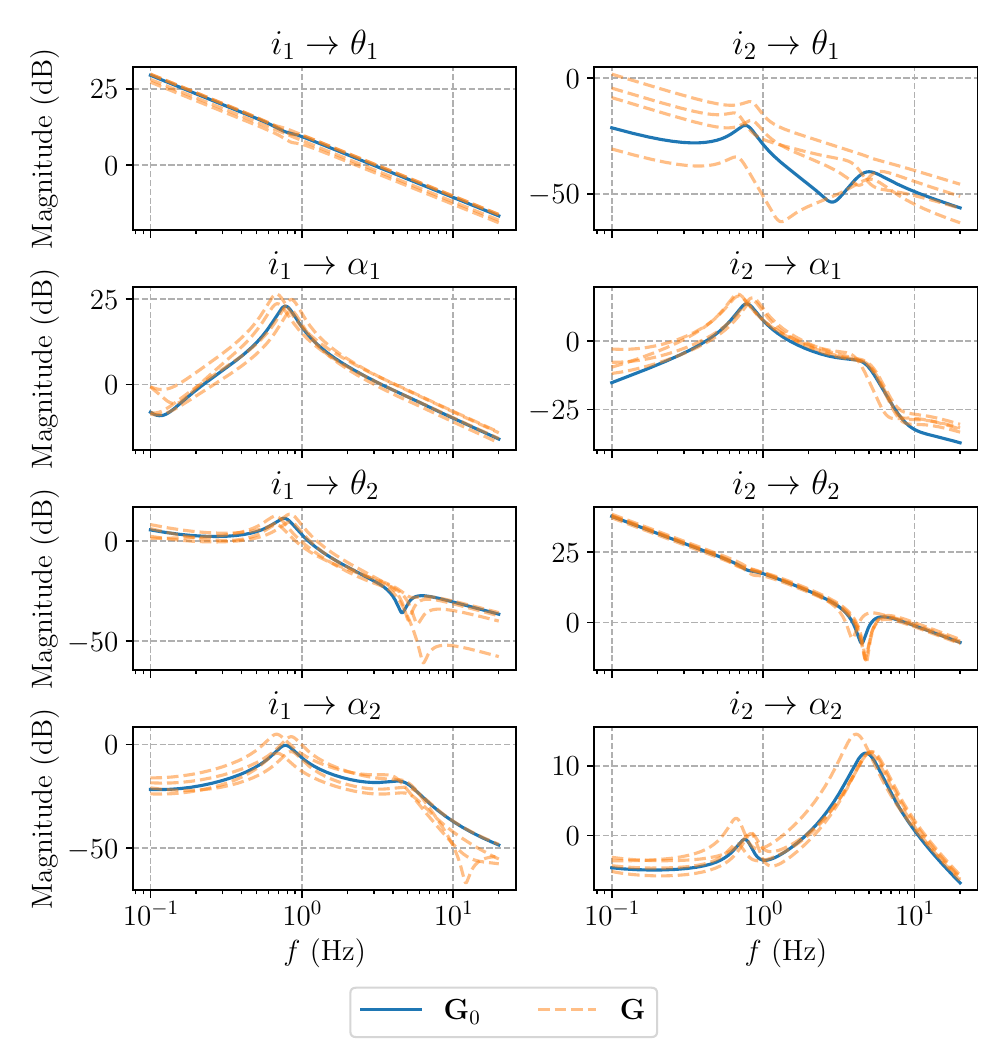}
  \caption{Magnitude response of the identified process models $\mbf{G}(s)$ in
  the variation set and the nominal model $\mbf{G}_{0}(s)$.}
  \vspace{-15pt}
  \label{fig:mag_nom_offnom}
\end{figure}

A process model from the inputs \eqref{eq:inputs} to the states
\eqref{eq:states} of the flexible manipulator in the linear regime is
determined for each joint stiffness configuration using frequency-domain system
ID \cite{pintelonSystemIdentificationFrequency2012}.
For each configuration, $20$ datasets are collected over a duration of $81.92
\, (\si{s})$ at a sampling frequency of $200 \, (\si{Hz})$.
The current inputs applied to the system are odd random-phase multisines
\cite{pintelonSystemIdentificationFrequency2012}.
The frequency response matrices are computed using discrete Fourier transforms
of the input and output signals.
Then, state-space models are estimated using \texttt{MATLAB}'s \textit{System
Identification Toolbox} \cite{SystemIdentificationToolbox}.
The nominal model is obtained using the same system ID procedure with the
datasets from all the spring configurations.
Fig.~\ref{fig:mag_nom_offnom} shows the magnitude response of the identified
process models $\mbf{G}(s)$ and the nominal model $\mbf{G}_{0}(s)$.
The joint stiffness variation affects the manipulators' natural frequencies as
the peaks on the magnitude plots are shifted across the various configurations.
The measurement model is identical for all joint stiffness configurations and
is given by the constant matrix
\begin{align}
  \mbf{C} = 
  \begin{bmatrix}
    1 & 0 & 0 & 0 \\
    0 & 0 & 1 & 0 \\
  \end{bmatrix}.
  \label{eq:measurement_model_fjr}
\end{align}

\begin{figure}[t]
  \centering
  \includegraphics[width=1.0\linewidth]{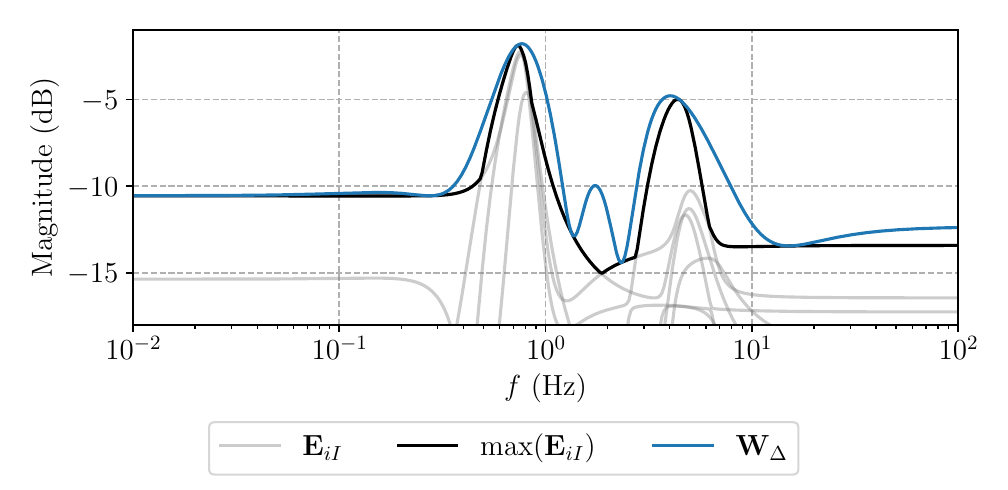}
  \caption{Magnitude response of the inverse multiplicative input uncertainty
  model characterization.}
  \vspace{-11pt}
  \label{fig:uncertainty_residuals_weight}
\end{figure}

Six common unstructured uncertainty models are evaluated for the uncertainty
characterization \cite[\S~8.2.3]{skogestadMultivariableFeedbackControl2005}.
The inverse multiplicative input uncertainty model is selected as it yields 
the smallest peaks in residual singular values, which implies less conservatism
in the uncertainty characterization.
The inverse multiplicative input uncertainty model is given by
\begin{align}
  \mbf{G}(s) = \mbf{G}_{0}(s) (\eye - \mbf{E}_{iI}(s))\inv,
  \label{eq:inverse_input_multiplicative_uncertainty}
\end{align}
where $\mbf{E}_{iI}(s) = \mbf{W}_{\Delta}(s) \mbs{\Delta}(s)$.
Fig.~\ref{fig:uncertainty_residuals_weight} shows the magnitude response of
the residuals $\mbf{E}_{iI}(j\omega)$ computed from $\mbf{G}(s)$
and $\mbf{G}_{0}(s)$.
The residuals are the largest around the natural frequencies of the flexible
joints as they are being perturbed by the variation in the stiffnesses.
The residuals do not increase with frequency due to unmodeled dynamics as is
typical in uncertainty characterizations because, in the present problem, the
uncertainty describes the variation between the models used in the observer and
the nominal model, all of which have the same structure arising from the
common system identification procedure.

The uncertainty weight $\mbf{W}_{\Delta}(s)$ is obtained by fitting a stable
and non-minimum phase transfer function whose magnitude overbounds the maximum
residuals $\max(\mbf{E}_{iI})(j\omega)$ using a constrained log-Chebyshev
method \cite{wuFIRFilterDesign1999}.
The fit of $\mbf{W}_{\Delta}(s)$ introduces conservatism in the uncertainty
characterization as it does not tightly overbound $\max(\mbf{E}_{iI})(j\omega)$
at all frequencies.
However, it accurately captures the peaks of the residuals, which is of greater
importance.

\begin{figure}[t]
  \centering
  \vspace{5pt}
  \includegraphics[width=0.9\linewidth]{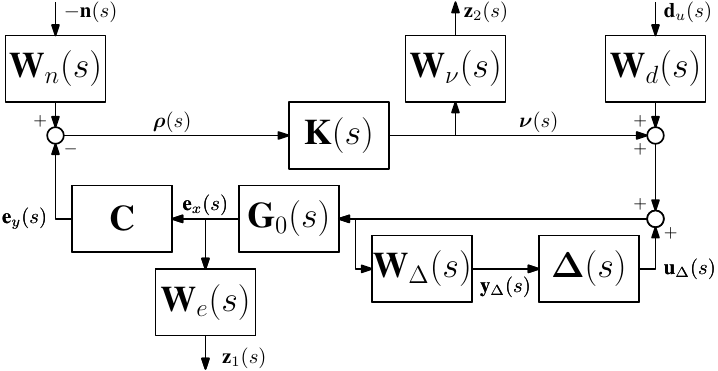}
  \caption{Robust observer generalized plant diagram.}
  \vspace{-10pt}
  \label{fig:generalized_plant_diagram}
\end{figure}

\begin{figure}[t]
  \centering
  \includegraphics[width=1.0\linewidth]{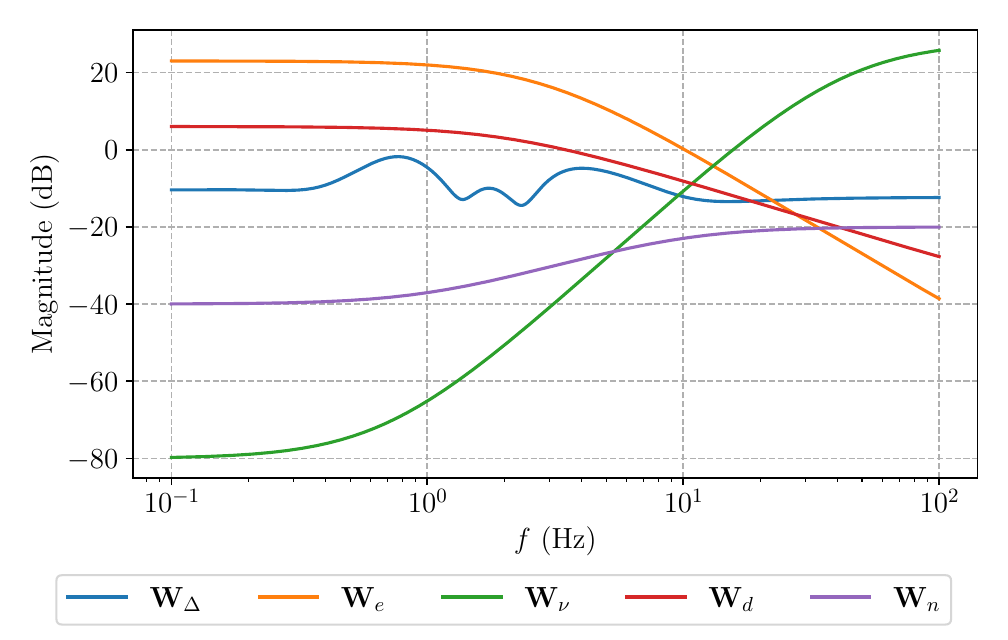}
  \caption{Magnitude response of the generalized plant weighting functions.}
  \vspace{-10pt}
  \label{fig:generalized_plant_weights}
\end{figure}

The generalized plant $\mbf{P}(s)$ is created by augmenting the input-output
observer error dynamics with filters that encode the performance requirements.
The block diagram of the augmented error dynamics is shown in
Fig.~\ref{fig:generalized_plant_diagram} and the magnitude response of the
weighting filters is shown in Fig. ~\ref{fig:generalized_plant_weights}.
The exogenous inputs and performance outputs are given by
\begin{align}
  \mbf{w}(s) &=
  \begin{bmatrix}
    \mbf{d}_{u}(s)^{\trans} & -\mbf{n}(s)^{\trans}
  \end{bmatrix}^{\trans},
  \label{eq:exogenous_inputs} \\
  \mbf{z}(s) &=
  \begin{bmatrix}
    \mbf{z}_{1}(s)^{\trans} & \mbf{z}_{2}(s)^{\trans}
  \end{bmatrix}^{\trans},
  \label{eq:performance_outputs}
\end{align}
respectively.
The input disturbance $\mbf{d}_{u}(s)$ and sensor noise $\mbf{n}(s)$ are
filtered by $\mbf{W}_{d}(s)$ and $\mbf{W}_{n}(s)$, respectively, to encode
the bandwidths over which they are expected to be significant.
The performance output $\mbf{z}_{1}(s)$ is the state estimate
error $\mbf{e}_{x}(s)$ low-pass filtered by $\mbf{W}_{e}(s)$ to penalize
estimation errors over the dominant frequency range of the dynamics.
The performance signal $\mbf{z}_{2}(s)$ is the input correction
$\mbs{\nu}(s)$ high-pass filtered by $\mbf{W}_{\nu}(s)$ to suppress any
high-frequency noise in the state estimate.

\begin{figure}[t]
  \centering
  \includegraphics[width=1.0\linewidth]{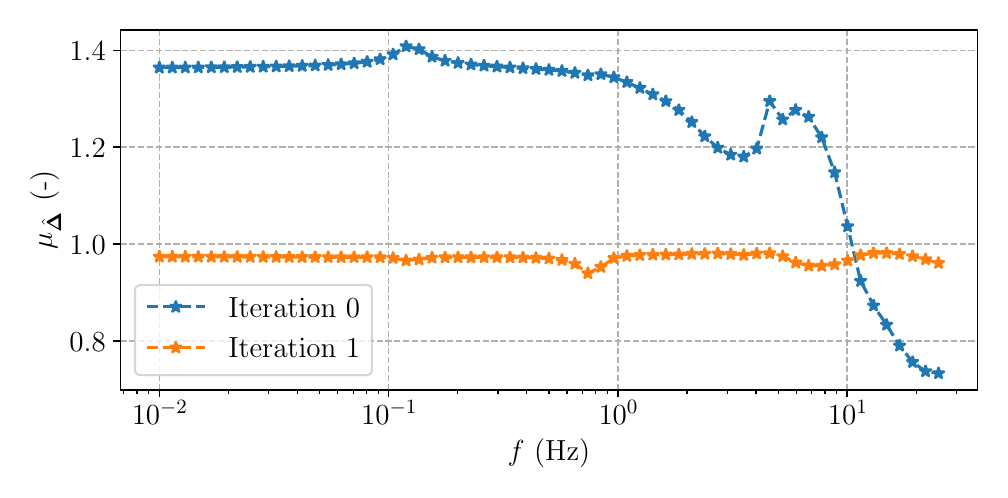}
  \caption{Structured singular value upper bound across iterations of
  $DK$-iteration.}
  \vspace{-10pt}
  \label{fig:ssv_iterations}
\end{figure}

The correction filter $\mbf{K}(s)$ is synthesized using $DK$-iteration
implemented in the \texttt{python} package \texttt{dkpy}
\cite{dahdahDecargroupDkpy2024}.
The SSV is evaluated on a logarithmically spaced frequency grid with $61$
points from $0.01 \, (\si{Hz})$ to $25 \, (\si{Hz})$.
The frequency grid is selected as it captures the dominant system dynamics with
sufficient resolution, particularly the natural frequencies of the flexible
joints at roughly $0.7 \, (\si{Hz})$ and $5 \, (\si{Hz})$ seen in
Fig.~\ref{fig:mag_nom_offnom}.
Fig.~\ref{fig:ssv_iterations} shows the SSV upper bound across the steps of
$DK$-iteration.
In the initial iteration, the SSV is greater than $1$ at nearly all frequencies
in the grid with a peak around $1.4$.
Upon the next iteration where an $8$th order fit of $\mbf{D}(s)$ is used, the
SSV is below $1$ at all the grid frequencies indicating that the robust
performance criterion is achieved.

\begin{figure}[t]
  \centering
  \includegraphics[width=1.0\linewidth]{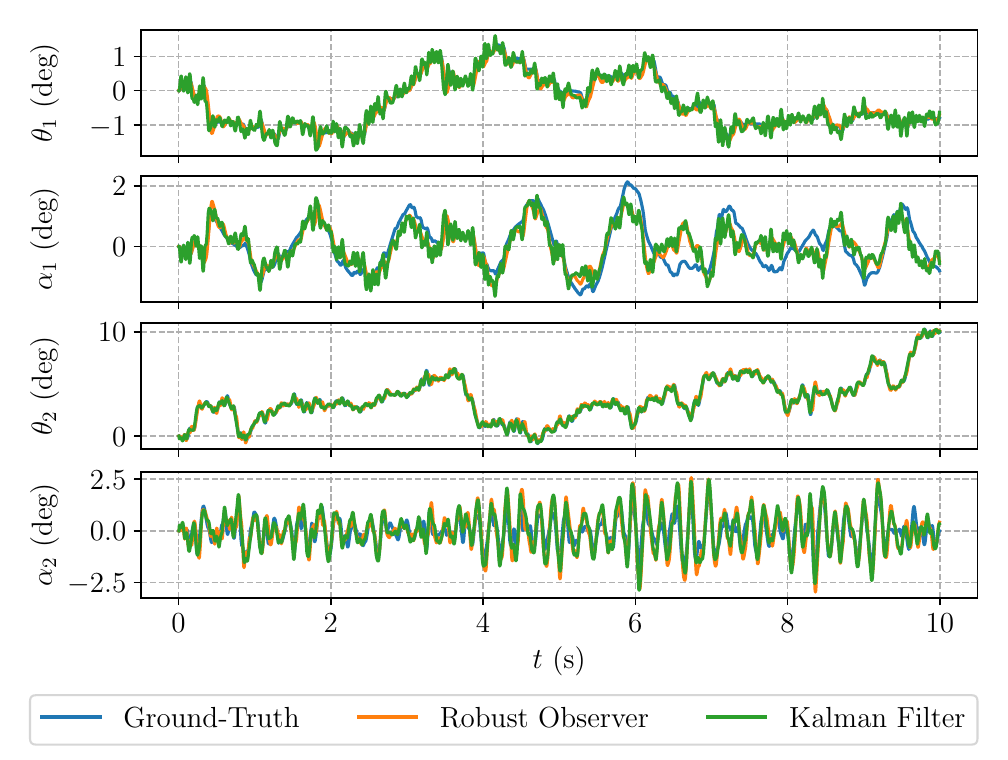}
  \caption{Estimated angular positions for the robust state observer and the
  tailored steady-state Kalman filter for the first $10 \, (\si{s})$ of an
  example joint stiffness configuration.}
  \vspace{-10pt}
  \label{fig:estimator_results_state_estimation}
\end{figure}

The robust observer estimation is demonstrated on a dataset of the flexible
joint manipulator.
The robust observer is compared against a steady-state Kalman filter where the
Kalman gain is designed for each varied system, which is equivalent to the
procedure in Fig.~\ref{fig:overview_diagram_tailored}.
The Kalman gain is designed using process and measurement noise covariance
values that correspond to the DC gain of the disturbance filter
$\mbf{W}_{d}(s)$ and noise filter $\mbf{W}_{n}(s)$, respectively.
Fig.~\ref{fig:estimator_results_state_estimation} shows the angular position
estimates of the robust observer and tailored Kalman filter with the
ground-truth data for the first $10 \, (\si{s})$ of an example dataset.
The robust observer and Kalman filter track the ground-truth angular positions
well, with the exception being for $\alpha_{1}$.
This is likely due to the fact that the shoulder joint is not sufficiently
excited to overcome friction effects as both the robust observer and Kalman
filter exhibit a similar degree of error in estimating this angle.

\begin{figure}[t]
  \centering
  \includegraphics[width=1.0\linewidth]{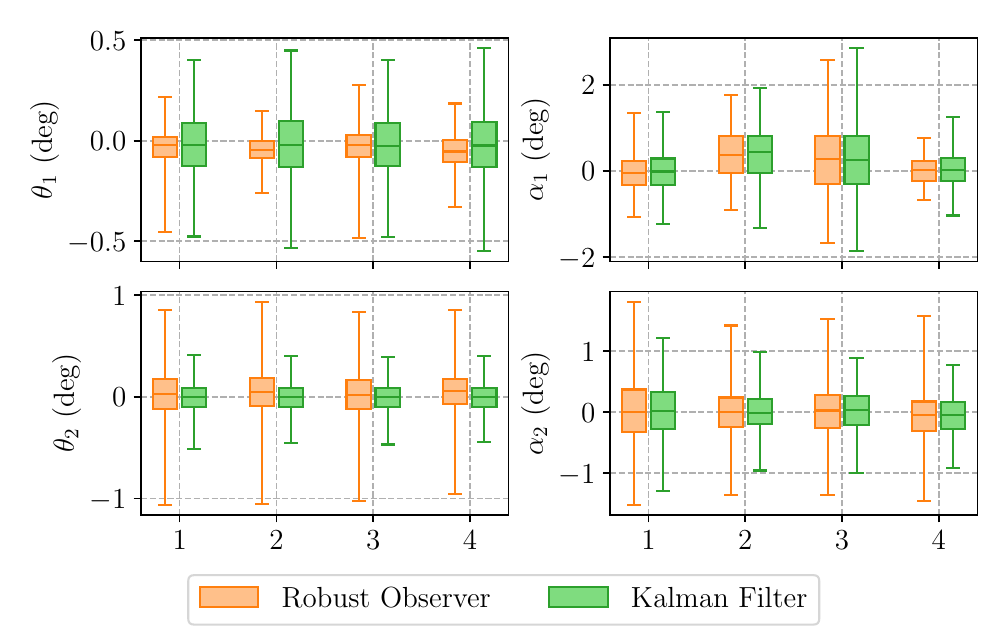}
  \caption{Box plot of the absolute error of the estimated angular positions
  for the robust state observer and the tailored steady-state Kalman filter
  across the $4$ joint stiffness configurations.}
  \vspace{-11pt}
  \label{fig:estimator_results_absolute_error_box_plot}
\end{figure}

Fig.~\ref{fig:estimator_results_absolute_error_box_plot} shows box plots of
the absolute error for each angular position estimate across the $4$ stiffness
configurations.
Both estimators have similar error distributions for $\theta_{1}$, although the
Kalman filter's peak errors are slightly larger.
Both state estimators have similar error metrics for $\alpha_{1}$, which is
slightly larger than the other angles likely due to the aforementioned issues
with the shoulder joint excitation.
The Kalman filter attains more accurate estimates for $\theta_{2}$ and
$\alpha_{2}$ as the peaks and the interquartile range of the absolute error are
smaller.
However, the difference is only on the order of a couple tenths of a degree
indicating that the two methods have similar performance.
Moreover, the larger absolute error seen in $\theta_{2}$ compared to
$\theta_{1}$ is likely due to the fact that $\theta_{2}$ attains larger angular
deflections. 
This demonstrates the effectiveness of the proposed observer design procedure,
which only requires the synthesis of a single correction filter, as it performs
comparably to a method that synthesizes of a correction gain for each system.

\section{Conclusion}
\label{sec:conclusion}

This paper proposes a methodology for synthesizing input-output observers for a
population of devices that exploit knowledge of the particular LTI system
dynamics.
The method requires the synthesis of a single correction filter that provides
guarantees on the estimation accuracy across the population.
The variation in the population is characterized, and a robust performance
problem is posed and solved using $DK$-iteration.
The design methodology is demonstrated on a flexible joint robotic manipulator
with variation in the joint stiffness.
It is shown that the proposed methodology provides comparable performance to a
standard design procedure while requiring the synthesis of only a single
correction filter.

\section*{Acknowledgment}

The authors would like to acknowledge Steven Dahdah for their discussions
on robust observer design.

\printbibliography



\end{document}